# Fusion research in a Deuterium-Tritium tokamak

Emilia R. Solano

*Laboratorio Nacional de Fusion, CIEMAT, Madrid, Spain*

*Writing as an independent expert, not representing institutional views.*

*Abstract:*

*The recent ITER re-baselining calls for new fusion-relevant research best carried out in a DT-capable tokamak device with similar technical choices. The present paper describes key issues that could be addressed in a Suitably Enhanced DT-capable Tokamak (SET), with tungsten plasma facing components, boronization systems, and 10 MW of ECRH, based on JET's characteristics and knowledgebase. We discuss hardware options, and show that fusion-relevant operational scenarios could be achieved. Notably, development, validation and testing of fusion and nuclear diagnostics, to be used in next generation devices, would require a DT-capable tokamak as described.*

## 1. Introduction

In this perspective article we consider the value to fusion development of research in a JET-like tokamak operating Deuterium+Tritium (DT) plasmas, as we see it today. We discuss how such research could inform choices to de-risk next-step fusion developments. We do not advocate building a new copy of JET. Instead, we describe how much could be gained by refurbishing and upgrading it.

In terms of hardware, the recent ITER re-baselining [1],[2],[3],[4] led to a change from the previously-planned Beryllium (Be) wall to a Tungsten (W) wall, and added a boronization system to deposit boron films. Additionally, re-baselined ITER proposes to increase its planned initial Electron Cyclotron Resonance Heating (ECRH) power, aiming at higher flexibility for its experimental programme, reduction of risks (such as compensating possible deleterious effects of W radiation on the plasma) and early achievement of Q = 10 with low neutron fluence. For instance, EU-DEMO already planned full W coverage [5] and pure ECRH [6] for auxiliary heating. The re-baselined ITER is expected to provide useful information for reactor designs predicated on the same choices.

Re-baselining has significantly modified the ITER Research Plan. We don't wish to distract the reader from the purpose of this article with an extensive description of the ITER re-baseline motivation and studies, recommending references [1-4] instead. But because the ITER first plasma is not expected in 2026 any longer, we think it is important the reader be aware of the new expected ITER timeline and the objectives of its early phases, summarized in Table 1. This re-baselined ITER Timeline motivates us to explore potential advances in tokamak physics and technology, particularly with DT plasmas, that could be pursued between now and 2042, when such plasmas are scheduled for development in ITER.

| **Operation Phases** | **Approx. Duration** | **Hardware, fuels** | **Aims** |
|---|---|---|---|
| Start of Research Operation | 2035-2036 | Water-cooled W divertor<br>**Un-cooled W 1st wall**<br>ECRH 40 MW<br>ICRH 10 MW<br>Hydrogen (H)<br>Deuterium (D) | 15 MA Hydrogen L-mode<br>7.5 MA, 2.65T Deuterium H-mode<br>Test plasma control, Glow Discharge Cleaning (GDC), boronization, Disruption Mitigation System (DMS), inventory control, ELM control. |
| DT-1<br><br>Fusion Power Operation-1 FPO-1 | ~2040 | **Tritium plant**<br>**Water-cooled W 1st wall**<br>ECRH: 60-67 MW<br>**H-NBI 33 MW**<br>ICRH 10-20 MW<br>H, H+D, H+T, T?<br><br>Low neutron/activation, human access possible | Repeat above tests with new systems<br><br>Aim for 50 s plasmas<br>Develop Hydrogen H-modes<br>H+T mixed plasmas (L and H-modes)<br>Optimise GDC, boronization, ICWC, ECWC<br>Fuel retention, inventory control<br>Run-away suppression tests with DMS |
| DT-1<br>FPO-2 | ~2042 | DT plasmas<br>Change to D-NBI? | Re-tune plasma control and protection Optimize DD and DT plasmas<br>Aim for $P_{fus}$=100 MW, $Q \geq 1$, 50 s |

Table 1: Simplified summary of expected ITER Operation Phases, based on Fig. 3.2 of Ref. **Error! Reference source not found.**.

Based on the ITER re-baselining choices and earlier DT experiences at JET, we consider the following questions, of interest for the next fusion steps:

1) how does full tungsten (W) coverage affect operation and fusion capabilities? Is boronization a practical solution to off-set the possible difficulties associated to the W wall and limiters? These issues will be addressed in Section 2.

2) can additional ECRH help control tungsten penetration into the plasma? To be addressed in Section 3. How much would ECRH facilitate tokamak operation? How, and how well, does electron heating heat ions to create and/or maintain fusion plasmas? See Section 4.

3) what other useful physics studies could be carried out in a DT-capable tokamak, before DT operation begins in next-step devices? These are explored in Section 5.

4) Are there fusion and nuclear diagnostics and measurement techniques that can be developed and validated in advance of DT operation in future devices? See sections 6 and 7.

Rather than describe what could be done to address the above questions in the many future DT devices that are presently being discussed, proposed, designed and built, we focus on what could be achieved with a DT-capable Suitably Enhanced Tokamak (SET), with results to be delivered before ITER commences its DT operations (~2042). For concreteness, SET is based on JET: R= 3 m, high plasma shaping capability, maximum current ~4 MA, maximum toroidal magnetic field 4 T, fully and (so far) uniquely DT-capable, with ~30 MW of Neutral Beam Injection (NBI) heating and ~6 MW of Ion Cyclotron Resonance Frequency Heating (ICRH). Additionally, to explore the impact of W limiters and walls, and of electron heating, on DT operations, SET would have Tungsten (W) plasma facing components, a boronization system and at least 10 MW of ECRH. Some additional modifications and upgrades could provide unique information on fusion diagnostics and controls systems.

Why start with JET? JET is currently the only Deuterium+Tritium (DT)-capable tokamak worldwide, within a fully integrated plant that was designed and built to handle these fusion fuels, and with fusion and Tritium-relevant diagnostics. Since start of operation in 1983, JET has played a crucial role in the development of nuclear fusion as a viable source of energy. In 1991, it became the first tokamak to produce controlled fusion power using a DT fuel mix [7]. In 1997, during the first full DT experimental campaign (DTE1) a fusion record was set, producing 22 MJ of fusion energy in 6 seconds [8], [9]. New Tritium and DT experiments, with a Be wall and a W divertor, were carried out in 2021 in an experimental campaign named DTE2 [10], [11]. Most recently another DT campaign, DTE3, took place in 2023 [12]. Of the many recent records, we could mention that 69 MJ of fusion energy were produced with Tritium-rich mixed DT plasmas [12],[13]. It is a world record 18× higher than the 2022 fusion energy record produced by the National Ignition Facility [14], but not so well advertised. These experiments provided invaluable data on plasma behaviour [17], heating and fuelling, energy confinement, thermonuclear alpha heating, fusion ash exhaust, fusion measurement techniques and the interactions between plasma and the reactor walls. Additionally, neutronics and nuclear activation measurements performed during these DT campaigns [15] are informing necessary developments in nuclear modelling, technology, protection and diagnostic design, required for safe operation of fusion reactors.

We must clarify that obtaining fusion records is not (and should not be) an objective *per se* of SET: the records are but one measure of proximity to the conditions to be expected in a fusion reactor, many others can and should be considered. Research in plasmas without Tritium is much easier and cheaper, and it is valuable to improve and validate understanding and to develop plasma scenarios and fusion technology that we think would be useful in DT plasmas. Testing those developments in DT contributes significantly to validate the theory-based predictions of fusion performance in a tokamak reactor and to reduce uncertainties and risks in the path to develop fusion. We will discuss this is some detail in what follows.

It would be very interesting to learn if a W wall only introduces minor additional changes to the plasma, thereby validating results already obtained in JET-ILW with a Be wall and a W divertor, or if the new conditions lead to sufficiently different results to warrant re-evaluation of earlier conclusions. Changes in SET results due to the added W PFCs, boronization and ECRH could imply modified expectations for future devices, possibly expanding the operating space previously explored in JET-ILW plasmas beyond the achievements of the experimental campaigns DTE2 and DTE3.

In what follows we explore key fusion-related challenges that a SET DT-capable tokamak could address. After suitable plasmas are developed, we find the potential advancement of fusion and nuclear diagnostics in high-performance DT plasmas as one of SET's most compelling benefits. Equally important would be its role in preserving and transmitting essential DT and fusion expertise to future generations.

The article is organised as follows: in section 2 we discuss the expected effects of W plasma facing components (PFCs) that could be investigated at SET, in section 3 we present the benefits of the ECRH application for plasma performance and access to a reactor-relevant parameter space, discussing in section 4 the joint impact of both W and ECRH on operating scenarios, including basic transport modelling of some target operating regimes. Possible dedicated isotope studies are described in Section 5. Fusion diagnostics are described in section 6, nuclear issues on section 7. Discussion is presented in Section 8. Appendices at the end add relevant information.

## 2. Effects of Full Tungsten Plasma Facing Components (W-PFCs) and Boronization

A description of the candidate wall composition for SET, as similar as possible to the choice in re-baselined ITER, is presented in Appendix A, and illustrated in Fig. 1. Basically, all limiter and divertor surfaces would be W or W-coated, while various alternatives are possible for remote areas of the wall: W-coated graphite tiles, steel, Eurofer, Inconel, etc.….

There is a concern that a W wall will constitute an additional W and radiation source affecting plasma performance in ITER [2], in excess of what was estimated with only a W divertor [18]. Expected W radiation in ITER has been modelled based on AUG data [19]: the dominant tungsten ion stage radiating in the confined plasma will change with temperature, going from potentially Ni-like $W^{46+}$ for $T_e$_core~4 keV to perhaps Ar-like $W^{56+}$ for T~10 keV plasmas, to C-like $W^{68+}$ for the hottest plasmas. Radiation sets limits to the W concentrations that can be allowed in the core plasma. However the cooling efficiency of tungsten decreases by a factor of ~1.5 over this temperature range. Notably, substantial radiation below 1-2 keV remains poorly understood and quantified, and it has important consequences on Baseline plasmas. W influx can be influenced by PFC composition, boronization, strike point choice, shape, ELMs and core MHD. Understanding this is a rather complicated multi-scale modelling problem. Therefore developing pragmatic operational approaches is essential. The observed impact of W on JET-ILW operating regimes will be discussed in section 4.

Although the total wall W flux of diverted plasmas may be sensibly smaller compared to the one from the divertor region, depending on plasma conditions the Scrape off Layer (SOL) screening efficiency could be smaller and the penetration of W to the plasma core larger. Additionally, the contribution of wall W sputtering due to seeding impurities such as Ne may put a limit to the maximum injected radiator element [20].

To mitigate the risks posed by the replacement of the Beryllium wall with a W main wall, the ITER re-baseline includes the installation of a conventional diborane ($B_2H_6$ or $B_2D_6$) boronization system. This will provide Oxygen gettering and help reduce W contamination of the plasma. The equivalent system would be installed in SET to document the need for and effects of Boronization: how long would its benefits last, characterise the formation and loss of Boron (B) coatings, B transport to remote areas, dust formation and fuel an impurity retention in W and in B layers. Most of these plasma-surface interaction studies would take place in SET early on, in Deuterium, but later on SET would eventually enable detailed characterisation of possible isotope effects in plasma wall interactions by studying plasmas with increasing amounts of Tritium. For instance, it would be important to quantify the production of B dust from the deposited layers with loose adherence to the metal surfaces, contributing to Tritium retention.

Starting from an initially pristine W wall would allow tests of low density breakdown at ITER-relevant low voltages. Next, plasma start-up in ITER begins with the plasma limited on the inner wall. After breakdown, ITER will have a long (~10s) limiter phase: it is expected that ECRH-assist will be needed to sustain the plasma while the current ramps up. But the density is still low so ECRH absorption (X3 in this case), as well as stray field radiation, is a problem that has to be better understood and quantified: is there a limit to the maximum ECRH power that can be used, due to stray radiation effects? Joint experiments in several machines are planned to study this burn-through problem, but none can do it with plasma currents above 1 MA like SET could, both in D and DT.

Beyond the plasma-wall interactions and scrape-off layer physics, the presence of high Z impurities in the confined plasma has long been a concern in the fusion community. Very early transport studies already described fast inward convection of high Z impurities due to electron density gradients [21]. Recent studies ([22],[23] and references therein) showed that the inward W convection proportional to $Z \cdot \nabla n_e$ can be balanced at least in part by outward convection proportional to $\nabla T_i/2$. ELMs and core MHD can also influence W behaviour, flushing W from the edge or moving it inwards. The overall balance of W flushing, penetration and/or screening across the pedestal is affected by plasma rotation and associated centrifugal effects, MHD and turbulent outward transport. This topic will be addressed in Section 4.

ECRH in SET can provide the central heating necessary to drive W out, while boronization could contribute to reduce the W source. Optimised solutions could be developed in SET, also taking into account isotope effects: would Tritium plasmas lead to more W sputtering and require more frequent boronization, leading to increased retention? Would isotope-dependent ELM behaviour lead to a different fuelling optimisation to keep W flushing? It would be very interesting to evaluate how and when ECRH affects W penetration and radiation, even before new W PFCs are installed in SET, to distinguish the divertor W source from the eventual full W coverage.

Next let us consider the possible contribution of SET to fuel inventory studies. As demonstrated in 2023 on JET [24], in-situ Laser-Induced Desorption-quadrupole Mass Spectrometry (LID-QMS) can be used for direct measurements of the fuel inventory of plasma facing components without retrieving them from the fusion device. At SET, LID-QMS could provide D and T inventory monitoring, document the Deuterium and Tritium co/deposition and/or retention in between pulses, as well as evaluate wall cleaning techniques such as Ion-Cyclotron Wall Conditioning (ICWC) [25] in ITER-relevant configurations with long connection lengths. The results can have important consequences for the management of Tritium budgets not only in SET itself, but also in ITER, where the LID-QMS diagnostic is foreseen as Tritium deposition monitor. Specially programmed experiments in SET with varying Tritium concentrations would provide complementary Tritium retention measurements from specific types of plasmas and wall conditions.

Lessons learned from gas-balance experiments at JET indicate it would be important to improve dedicated in-situ gas instrumentation in SET to develop reliable techniques for Tritium inventory quantification [26].

W is also likely to affect plasma disruptivity (often caused by W radiation itself, see section 4) and runaway production, and it may be necessary to re-optimise disruption and runaway control techniques [27], [28]. Installation of a 2nd Shattered Pellet Injection system for disruption mitigation could be considered, maybe even a Tritium compatible one.

Runaway electrons following a disruption in a high current plasma are one of the major concerns for tokamak reactors, and documenting run-away deposition and potential erosion depth in W tiles would be of interest to next step devices [2] . It might be possible to install a few single solid (uncooled) W first wall and/or limiter tiles in strategical locations, to study tile damage from disruptions and run-aways [29]. One could consider mimicking the ITER choice of attachment for W plaques, to assess their robustness against disruptions.

## 3. Electron Cyclotron Resonance System: impact on operation

In the ITER re-baseline, an increase in the power to be delivered by the ECRH system in the initial phase of operations was deemed necessary to counteract possible additional radiation in the plasma from the tungsten limiters and first wall, to increase operational versatility (such as control of sawteeth and neoclassical tearing modes) and to ensure Deuterium H-modes would be accessible during FPO-1.

In SET we consider two options for the ECRH system. Option 1 is well documented, while Option 2 is more ambitious and would need development. Additional information is presented in Appendix B.

Option 1 is to adopt the 2011 design of an ECRH system for JET [30][31][32], off the shelf. It is a system of 12 gyrotrons, delivering 10 MW to the plasma, using the ITER gyrotron frequency of 170 GHz [33], relying on 2nd harmonic X-mode for SET. Steerable mirrors are used to select deposition locations and current drive direction. Such a system is optimised for off-axis current drive and Neoclassical Tearing Mode stabilization for a tokamak toroidal field in the range 2.7-3.1 T. Alas, this system cannot deliver core heating at higher toroidal fields.

Option 2 aims to deliver ECRH at multiple fields. Given that STEP [34][35] and EU-DEMO [36] both consider using ECRH systems with hundreds of gyrotrons, with a variety of frequencies, it would be beneficial to the fusion programme to install at least some of these systems early on in SET. Options being considered are multi-purpose/multi-frequency gyrotrons, and tuneable-frequency gyrotrons. For instance, for EU-DEMO the gyrotron frequencies presently under consideration are 136, 170, 204 and 238 GHz, which would be resonant at 2.4, 3, 3.6 and 4.2 T respectively. In SET such a broad range of frequencies could ensure delivery of central heating for W control, as well as off-axis heating and current drive for operational flexibility in plasmas with a broad range of toroidal fields. The whole system requires careful design and development of the gyrotron themselves, wave-guides and windows. It would be an ambitious undertaking to do this in SET, but it would bring benefits to the fusion community at large, as it would enable early tests of technological options that need to be robust when implemented in next step devices.

The use of ECRH for break-down and current ramp up assist was already alluded to in Section 2. Additional issues that could be addressed with ECRH in SET include:

### 3.1 Central ECRH for core electron heating and W control

Operating conditions at JET were changed substantially due to the W divertor. It is expected the localised central heating from ECRH can contribute to keep the W out of the core, as done with localised Ion Cyclotron Resonance (direct and indirect) electron heating in JET-ILW H-mode plasmas [37], and as planned in re-baselined ITER. We relegate this discussion to section 4 on operating scenarios, since W behaviour and the applications of core electron heating are different in hybrid and baseline plasmas.

ECRH would allow studies of plasmas with dominant electron heating in a broad range of conditions.

### 3.2 ECRH and Electron Cyclotron Current Drive (ECCD) for MHD control

In high beta plasmas with positive magnetic shear, Neoclassical Tearing modes (NTM) can lead to reduced confinement. Since the early 2000's a variety of studies have shown that localised ECRH or ECCD can be used for NTM control.

Equally important, the ECRH system would also enable sawteeth pacing, which also relates to W contamination: a large sawtooth crash can allow peripheral W to penetrate into the core, leading to large core radiation and hollow electron temperature profiles, and eventually a disruption or a controlled plasma termination. Alternatively, an opportune sawtooth can expel W from the core. Control loops for sawtooth pacing and W expulsion could be set up with the ECRH system, reacting to hollowness measurements.

### 3.3 EC for Internal Transport Barrier creation and control

Localised ECRH and ECCD could be used to control Internal Transport Barriers (ITBs), raising ion temperatures and thermonuclear alpha particle population and thereby producing stronger burning plasma effects than in the experiments carried out so far on JET-ILW [38], [39],[40].

Hot ion ITB regimes provided the basis for plasmas especially designed to investigate alpha heating of electrons [39] and alpha-driven instabilities in DTE2 [38], using only NBI and obtaining at best ITBs with $T_{i0}$=12keV and $P_{fus}$=4MW. Long ago, ITB plasmas with optimised magnetic shear and $T_{i0}$=30 keV were produced in JET-C by early off-axis heating or current drive during the plasma current ramp-up phase [41] to “freeze-in” the q profile. Later application of high central heating in such plasmas lead to avoidance of sawteeth and formation of ITBs in D [41] and DT [42]. In JET it used to be possible to consistently produce ITB plasmas with the use of the Lower Hybrid Current Drive system, no longer available. In SET, ECRH and/or ECCD could fulfil that role.

Central heating of electrons without introducing ICRF-driven fast ions would be a critical new capability, restoring the ability to achieve high central $T_e$ before and during application of NBI. ECRH would allow less reliance on NBI power, and thereby operation at lower density, as well as longer slowing down times, increasing the alpha particle population. This type of scenario could be re-developed with either ECRH option, hopefully recovering the ITB plasmas with $T_{i0}$=30 keV that were observed in the JET-C campaigns with low density and early heating [41], in DT, and possibly with ECCD stabilised ITBs. It would prove useful for fusion diagnostics, to be described in section 6.

### 3.4 Combined heating: ECRH and fast ion populations from ICRH or NBI

Comparison of plasmas with different ratios of total heating delivered by the various systems would help elucidate transport channels and investigate the impact of rotation on transport.

In JET, the main heating source available is NBI, capable of injecting H, D, T and $^4$He ions. H-NBI is difficult in a positive ion system due to the lower neutralisation for H beams and hence a higher power load on the ion dumps. In 2014 a water leak occurred on one of these ion dumps within the one of the JET beamlines and this led to a limit being placed on maximum H-NBI power during the 2016 H campaign, and a general reluctance to use H-NBI subsequently. T-NBI was successfully deployed in DTE2 [43], successfully handling T containment issues [44] D-NBI is very robust and easy to condition, which is why it was the choice for DTE3.

In JET-ILW, ICRH was often used to deliver central electron heating for W control [37]. In SET, as the ECRH would deliver central electron heating, the ICRF system would be liberated for ion heating. In DT plasmas this can be achieved with combined $^3$He minority and 2$^{nd}$ harmonic Tritium schemes, as planned for ITER DT plasmas and already demonstrated at JET-ILW [45][46]. Alternatively, ICRF heating of D ions has been shown to significantly boost the net fusion reactivity, since both the thermalized D ions and the fast D-NBI ions are accelerated to energy ranges that are optimal for the DT reaction cross-section [47]. Development of 3-ion ICRF scenarios for heating bulk ions in DT plasmas already proven at JET [48] could continue in SET.

With higher $T_e$, produced by central ECRH, fast ion populations (from NBI, alphas, ICRH) would be larger because their energy content depends linearly on the longer slowing down time, which scales with electron temperature as $T_e^{3/2}$.

If central ECRH works well, keeping W out of the core plasma by central electron heating, the combination of NBI, ICRH and ECRH is expected to deliver plasmas with higher core $T_e > T_i$ at high densities (see Baseline section, 4.4) and/or with $T_e$ closer to $T_i$ in low density hybrid type plasmas (see Hybrid, 4.3). We present a discussion of such scenarios in the next section.

## 4. Expected SET operation scenarios

After the installation of the JET-ILW Be wall and W divertor, it took years for the JET team to optimise Deuterium plasmas to be used as references for DT experiments. Most of the operational difficulties were associated with W contamination of the plasmas and not always sufficient available central heating power to overcome the increased radiation. The W behaviour differed between the two main operating scenarios, hybrid and baseline. Hybrid plasmas have low current and density, high beta poloidal ($\beta_\theta$), typically $T_i > T_e$, and maximal fusion performance (see 4.1). Baseline plasmas have high current and density, low $\beta_\theta$, $T_i$=$T_e$ and typically maximise thermal fusion (see 4.2).

**4.1. Influence of W on operation scenarios:**

In low density plasmas low fuelling can lead to infrequent large ELMs, which in JET-ILW could cause W sputtering and introduce W into the plasma [49],[50]. Initially, steep edge density gradients can drive W inwards from the SOL towards the top of the density pedestal, while centrifugal forces (due to NBI-induced toroidal rotation) force the W to stay in the low field side, creating a crescent shaped toroid with high radiation. The development of this W "radiating crescent" has been documented [50], [51] and reported to be a transient phenomenon in hybrid plasmas in JET-ILW. It was found that low density hybrid operation with sufficient central heating had high ion pedestal temperatures that resulted in (at least partial) W screening: the W neoclassical drift is outward (see section 2.3 of [52] for details), slowing down core contamination. ELMs can both increase W influxes and flush W from the edge. Hybrid scenario optimisation had to take into account these various competing effects, as discussed in some detail in section 2 of reference [52]. Eventually a hybrid scenario was developed without the W radiating crescent, and with sufficiently slow core accumulation to allow high performance hybrid plasmas to be sustained for many confinement times. Even then, radiation of order 1/3 of NBI power (10 MW out of 30) was common in these plasmas.

In high density baseline plasmas, on the other hand, the W radiating crescent typically stayed stably at the outer edge [53], likely contributing to hold the pedestal edge electron temperature down. Especially in high current plasmas, any drop in heating power would lead to cessation of the ELM activity and associated ELM flushing [64], followed by an increase of density and edge radiated power and eventually a disruption caused by edge-cooling due to the W radiating crescent [51] [53], limiting the development of Baseline plasmas to 3.5 MA. JET pulse 99948, to be discussed in section 4.4.1, displays this behaviour. W radiation peaks at normalised poloidal flux of 0.8, where the electron temperature is initially of order 1.6 keV. When NBI power drops from 29 MW to 27 MW, the plasma radiation, initially at 15 MW, rises to 30 MW in 300 ms (much of it in the edge), leading to abrupt plasma termination.

How much boronization can contribute to mitigate or control W contamination in typical hybrid and baseline plasmas is one of the questions SET would need to address, in combination with the use of ECRH to drive W outwards by raising the temperature gradients.

Edge-core integrated DT **seeded baseline** plasma scenarios can display different behaviour. Neon-seeding induced detached divertor conditions that reduced W influxes while maintaining good core confinement [54] in conditions of high ion heating and rotation. Seeded scenarios have been developed up to 3.2 MA [55], with good confinement and no evidence of W contamination when sufficient heating power is available. Additional ECRH power would be helpful to their further development and understanding, especially to increase current and confinement. In particular, new studies could be carried out to optimise seeding of Neon also in the plasma ramp-up and ramp-down phases, to mimic ITER conditions, deploying real-time radiation control tools [56],[57].

Recently other JET scenarios focused on AUG-inspired DEMO-relevant regimes with very high-density plasmas, such as QCE [58] and X-point radiator [59], both obtained in DT in JET-ILW. The XPR often hovers near the L-H transition due to large radiation combined with power limitations. The QCE benefited from seeding, but run into technical limitations on total injected NBI power. These scenarios appear not to suffer from W contamination but their development also would benefit from the additional ECRH power, as it would enable H-mode studies at higher current and/or field, with plasmas simultaneously closer to fusion relevant pedestal values of Greenwald density, beta or collisionality, and with larger fractions of electron heating.

### 4.2. General considerations on operational regimes

The questions to be answered by SET in terms of operational regimes are self-evident: can they be recovered despite a W wall? Will boron layers survive high performance plasma pulses of different types? Will electron heating from ECRH be able to heat ions sufficiently to increase alpha particle production and take the SET plasmas closer to fusion-relevant conditions? Will the available ECRH power exceed the increased radiated power from the potentially larger W source, even at high density? Can the W radiating crescent be suppressed with on or off-axis ECRH, leading to hotter pedestals? How do the transport mechanisms in equilibrated $T_i$=$T_e$ plasmas compare to hot ion or hot electron regimes?

Note that in a fusion reactor, high ion temperature ($T_i$) must be obtained to start the fusion reactions that will heat the plasma. As described in [60], an initial period of low plasma density, in which the plasma accesses the H-mode regime and the alpha heating power increases, is necessary after the start of the additional heating, followed by a slow density rise. The start of Baseline DT shots in next step devices is likely to be initially a hot ion regime, with a density ramp taking them towards electron-ion thermally equilibrated Baseline plasmas. SET would enable dedicated studies of such delicate transient phases in DT plasmas. One might even mimic what might be possible in a fusion reactor with application of ion heating (ICRH or, less likely, NBI) to jump start fusion at the beginning of the pulse, and leaving only ECRH for fusion control once high density leads to e-ion thermalisation.

Modelling attempts to answer many of these questions for ITER. SET experimental tests can validate predictions, and experimentally characterise boundary conditions where models fail.

In the next two sub-sections we consider the Hybrid and Baseline scenarios in particular, as they deliver the highest fusion performance. We include predictive modelling studies to evaluate the benefits ECRH brings in both cases. Note that in what follows we only discuss core transport, with assumed pedestal values based on the JET-ILW experimental references. We can't predict expected changes in pedestal, ELMs or SOL that may occur due to the W wall. It is possible that by excluding W from the edge, ECRH might allow hotter pedestals. If this were the case, expectations of high-performance DT plasmas could be exceeded.

High performance DT fusion plasmas are expected to deliver the fusion alpha particles, alpha heating and 14 MeV neutrons needed to test, qualify and validate fusion and nuclear diagnostics, described in sections 6 and 7.

**4.3. Hybrid scenarios:**

Hybrid scenarios in D and DT with JET-ILW were developed to maximize fusion output in DT. The optimal DT plasmas achieved, leading to fusion energy records, had moderate (peaked) electron density $n_e$, low collisionality, high $\beta$, hot ions, central NBI power deposition, high $q_{95}$, low q-shear and moderate ELMs [52]. In SET, judicious use of combined heating might enable access to higher ion temperature ($T_i$) in core and pedestal, higher $\beta$ regimes with higher fast ion energy content, and possibly turbulence stabilisation.

W behaviour can also lead to unwanted plasma termination. Even though low-density plasmas can be made to partially screen W from the plasma, it has been observed that transient drops in heating can lead to penetration of W to the core, in some cases increasing core radiation to hollow out the $T_e$ profile and produce a disruption. Adequate plasma control strategies were developed to terminate such plasmas as soon as $T_e$ hollowness was detected [52].

### 4.3.1 Predictive core modelling of Tritium-rich Hybrid plasma:

Record non-thermal fusion energy production was achieved in JET-ILW in Tritium-rich hybrid plasmas. They were developed with large fast ion populations created by injecting D-NBI into Tritium plasmas and direct fundamental Deuterium ICRF heating with 29 MHz [47], maximizing the non-thermal reactivity [13]. In DTE3 the fusion record of 69 MJ of fusion power was produced with these techniques [12].

We begin with predictive core modelling of a plasma based on the fusion energy Tritium-rich record JET Hybrid pulse 104522. It had 80% Tritium, 20% Deuterium, 2.5 MA current, 3.9T toroidal field, 30 MW of D-NBI and 5-6 MW of on-axis RF power, reaching $T_e$=8 keV, with $T_i>T_e$ and produced about 13.5 MW of fusion power [12]. Temperature profiles near 9.5 s (a steady phase) are shown in blue in Fig. 2.

First, diffusion coefficients were estimated from the profiles of the reference pulse. A set of coupled Fokker-Planck equations is subsequently solved to shed light on the cross-talk of the various interacting and generally non-Maxwellian ion populations. Details on the numerical codes and procedures are described in [61], [62], [63]. Then the coupled energy transport equations were solved to predict the behaviour of this initial plasma when an additional 10 MW of central electron heating is added (to simulate ECRH), assuming unchanged transport coefficients and density profiles. The estimated temperature profiles with the added “ECRH” are shown in red in Fig. 2. As expected, adding 10 MW of core electron power primarily raises $T_e$, from 8 to 18 keV, but $T_i$ also increases from to 11 to 16 keV, which in turn results in an increase of fusion power from 13.5 to 19 MW. The “ECRH” has taken the reference hot-ion plasma to a situation with $T_e>T_i$, more representative of what is to be expected in self-sustaining fusion plasmas, and therefore much more interesting. If ECRH and the high $T_e$ succeed in excluding W from the plasma core, even higher performance could be expected.

As this was a high field pulse, 3.9 T, high frequency gyrotrons would be required to deliver central heating: the option 2 described in Sect 3 and Appendix B. Alternatively, a similar scenario would need to be developed at $B_{tor}$=3 T, using $^3$He ion-heating ICRH scenarios.

### 4.4 Baseline plasmas:

We turn now to high current, high density plasmas. Future tokamak burning fusion reactors will rely on plasmas as close as possible to thermal equilibrium, thus minimising the risk of fast particle populations triggering undesirable and often ill-controlled phenomena such as MHD activity, while guaranteeing abundant production of neutrons. Baseline plasmas were developed in JET-ILW to maximise thermal fusion output [53], and they have included dedicated DT experiments at JET in 2021 and 2023.

The fusion-relevant 50/50 DT isotopic mixture can be used to mimic baseline reactor plasmas as closely as possible. The high operating density of baseline plasmas leads to high equipartition efficiency: the power delivered to the electrons would increase $T_{ion}$ on a timescale well shorter than the typical high-performance duration of several seconds, boosting the neutron rate. The equipartition time for baseline plasmas is of the order of 1 second, as illustrated in Fig. 3.

The high densities pose the practical problem that the ~110 keV NBI deposits energy and momentum almost everywhere in the plasma, with a significant fraction in the outer layers of the plasma. The W radiating crescent described in section 2 typically stayed stably at the outer edge (see Fig. 3 of reference [53]), possibly contributing to hold the pedestal edge electron temperature down. But any drop in heating power would lead to cessation of the ELM activity and associated ELM flushing [64], followed by an increase of density and edge radiated power and eventually a disruption caused by edge-cooling due to the W radiating crescent [51]. This limited the development of high density baseline plasmas to 3.5 MA. ECRH would be beneficial in the first place to ensure sufficient heating is available, but also to extend the SET operational baseline regime to high current and density more reliably, both in D and DT. As will show. in SET ECRH is expected to both raise $T_e$ above $T_i$ and increase $T_i$, approaching more reactor relevant conditions.

We present two different predictions. In 4.4.1 we predict the behaviour of a 3.5 MA DT plasma, similarly to what was done in 4.3.1. In 4.4.2 a higher fidelity transport model that was validated for a 3 MA Deuterium pulse is used to predict changes to that plasma from added ECRH.

### 4.4.1 Predictive modelling of Baseline DT plasma:

The reference Baseline shot selected for this study was JET DT pulse 99948: 3.5 MA current, 3.3 T toroidal field, 29 MW of NBI and 3.5MW of on-axis RF power (fundamental H-minority at 48 MHz). It was a plasma with balanced D and T fuel content, with half D-NBI and half T-NBI. The adopted parameters are representative for the steady state (t=49.5s), which lasted about 2 s. As described earlier in section 4.1, this plasma has a W radiating crescent. When NBI power drops by less than 10% the W radiation at the edge increases beyond the heating power, leading to a disruption.

From experimental kinetic profiles, NBI and ICRH power deposition profiles are simulated generally as described in 4.3.1. Now the set of coupled Fokker-Planck equations is solved for a minority of Hydrogen, near-balanced D & T majority ion species and, finally, D & T beams, each of which is modelled accounting for the full, half and third energy subpopulation. Corresponding profiles are shown in blue in Figs. 4a and 4b. Note that the electron power deposition is peaking near the colder but still fairly high density edge, a result of the adopted beam heating.

Next, again assuming that transport coefficients and density profiles remain constant, only 3MW of additional central ECRH heating is applied, modifying the power deposition profiles, as shown in Fig. 4a. From that a predictive transport evaluation produces expected $T_e$ and $T_i$ profiles as a function of time, shown in red in Fig. 4b, compared with blue reference profiles. We see that the plasma evolves from $T_e$~$T_i$~8keV to $T_e$~10 keV, $T_i$~9 keV in 500 ms and $T_e$~11 keV, $T_i$~10 keV in 1 s, now reaching a regime with hotter electrons than ions, as expected in ITER. In the core the RF and the total power density is significantly higher than in the rest of the machine, with total power densities in excess of 0.6MW/$m^3$. Because of their differing mass, the deuterons collide with slowing down RF and NBI heated particles before the tritons do. Deuterons are also directly heated by the RF waves. Consequently, the D power density exceeds that of the T population.

From this estimate we conclude that SET, with only 3 MW of ECRH, already opens the door to fusion-relevant studies of Baseline DT plasmas with substantial electron heating and $T_e \simeq T_i$, hopefully excluding W from the plasma.

This scenario would need to be reoptimized for 3 T if Option 1 of ECRH is chosen, or could be optimised differently for best joint use of ECRH and ICRH in Option 2.

**4.4.2 Modelling the impact of ECRH on a Baseline Deuterium plasma:**

To support the analysis presented in 4.4.1 and to further assess the impact of ECRH on a plasma in SET with validated transport tools we performed a series of simulations based on previous modelling of an existing Deuterium JET-ILW pulse. The rationale and the advantage of choosing a D pulse as opposed to a DT plasma lies in the fact that the starting point is a thoroughly validated simulation of a well diagnosed experiment where most physics aspects are reasonably constrained. The scarcity of experimental data and the lack of a sustained high-performance plasma in DT mean that a DT plasma with the same level of in-depth modelling is not available yet.

The reference for this predictive simulation is JPN 96730, with 3 MA of plasma current, 2.8 T, 29 MW of NBI and 4.1 MW of on-axis RF power, D gas injection and 45 Hz D-pellets, a Deuterium JET pulse previously analysed in detail in [65]. Note that this plasma has lower current and field than the DT reference discussed in 4.4.1, and is therefore is automatically compatible with both Option 1 and Option 2 for the ECRH system.

Predictive modelling is carried out with JINTRAC [74], artificially adding to it 5 MW and 10 MW of ECRH heating respectively. In these simulations the electron density at the top of the pedestal was kept constant, but profiles and transport coefficients are evolved, aiming for higher fidelity than the simulation described in 4.4.1. The model adopted for these exploratory predictive runs is the Bohm/gyro-Bohm transport model and the ECRH power deposition profile was assumed an on-axis Gaussian with width 10% of the plasma minor radius.

The predicted results after 2 s are plotted in Fig. 5, where we show the converged electron $n_e$, $T_e$ and $T_i$ profiles. It is shown that with 5 MW of ECRH there are minor changes in these D-plasma profiles, while 10 MW are sufficient to switch to the desirable $T_e>T_i$ regime in the plasma core, the domain where the sign of the equipartition term in the energy balance is reversed, already in D plasmas. It is interesting to note that the addition of pure electron heating is predicted to flatten the electron density profile: we observe a decrease of the core electron density and an increase in both the $T_i$ and $T_e$, potentially increasing the thermonuclear power production in a matching DT plasma.

We believe that this modelling exercise is relevant also for DT plasmas as major differences on the effect of additional electron heating in going from pure D to DT are not be expected. In DT plasmas the electron-Tritium heat exchange would be slower than in Deuterium, but the power to the electrons would be the same. The experimental differences observed in the D and DT plasmas are largely associated with different ELM behaviour (hard to predict) and higher particle confinement when T is present, leading to increasing density and radiation, and shorter high performance duration. Integrated modelling accounting for density evolution remains a challenge. Experiments at SET, in D and DT, could provide valuable validation datasets for future models.

**4.5 Other operating scenarios for specific physics studies:**

A variety of other scenarios could be developed to address specific fusion issues. For instance:

- Conditions for confinement saturation and/or $T_i$ clamping [66] with central ECRH and ICRH at high density could be investigated in detail with the additional electron heating, and possibly compared with alpha heating in appropriate DT plasmas.
- Conversely, an investigation of the beneficial interaction between electron and ion thermal transport and instabilities driven by fast ions was initiated in JET, starting with fast $^3$He RF accelerated ions in low density Deuterium plasmas [67], [68] and continued in DT [69]. In these cases, turbulence was reduced in the presence of the increased fast ion population. As the balance between potentially detrimental effects ($T_i$ clamping) on fast ion confinement and turbulence suppression due to the instabilities remains largely unknown, the combination of ECRH, ICRH and NBI heating in the SET would provide unique opportunities for advancements in this area.
- Recently developed high density high $\beta_{poloidal}$ plasmas with turbulence suppression were observed both in DIII-D and EAST [70]. The exploration and development of such regimes in SET can contribute to pave the way towards steady state reactor scenarios.
- Long duration pulses of 30 or 60 s were obtained at JET in Deuterium in late 2023 [71],[72]. Such long pulses on SET would enable detailed characterisation of all transport channels in H-mode, especially using modulations of gas inlet, ECRH, ICRH, and/or NBI, in studies for D, T, H+T mixtures and DT for all transport channels [73]. They would also be a good target for nuclear measurements, such as water activation or single event effects (see section 7).
- Operation without a divertor cryo-pump in D and/or T could mimic ITER's relatively low pumping capacity (this is a size effect, with great impact on Tritium consumption) and enable detailed studies of particle transport by ELMs, and/or in no ELM regimes.
- Investigation of typical operating regimes with Tritium-poor mixtures with H or D in SET would inform ITER predictions for its FPO-1 phase.
- ITB scenarios for Energetic Particle studies [38], including alpha heating [39], as ITB plasmas would be easier to develop and maintain with the EC system, and fast ion populations should be higher high higher $T_e$ (see section 3.3).

In all cases, the additional ECRH power could allow more reliable operation, and enable experimental scaling studies, for instance, of confinement and scrape-off layer widths, at higher current, fields and/or densities.

## 5. Dedicated Isotope physics studies in SET:

Isotope studies, investigating how the plasma composition affects particle and energy confinement as well as ELM behaviour, are generally done in most devices comparing H and D, and sometimes $^4$He (it used to be a candidate for low activation phases of ITER). In SET, of course, uniquely relevant studies for next-step devices can include Tritium, Hydrogen-Tritium mixtures and Deuterium-Tritium mixtures. Often results are presented in terms of the plasma's mass A, or (in mixed plasmas) its effective mass, $A_{eff}$.

A complication that can sometimes be turned into a tool is that plasma behaviour in JET-ILW is known to depend strongly on plasma shape [75], [76], [77] specially near the X-point, and these changes may override sometimes subtle isotope effects. Typical plasma shapes used in isotope studies are shown in Fig. 6. The configuration with lowest L-H transition power threshold, ($P_{LH}$) is called Horizontal Target (HT): it has the inner strike in a vertical target and the outer one in the central almost horizontal divertor tile, with the highest lower triangularity. HT has the least pumping but lowest confinement once in H-mode. The Corner configuration, with strikes near the locations with optimum pumping, has high $P_{LH}$, lower $n_{e,min}$, and good confinement once in H-mode. The Vertical Target (VT) has both strikes on vertical targets: like Corner, it has high $P_{LH}$ and low $n_{e,min}$.

The effect of configuration on $P_{LH}$ confinement and Edge Localised Modes (ELMs) is not fully understood yet, and it may have strong implications for good H-mode access in next step devices. Various isotope studies have been carried out with different configuration choices, complicating interpretation. SET would provide an opportunity to unify some of the observations by designing experiments with matching configurations.

### 5.1 L-H studies

In both AUG [78][79] and JET it has been reported that the change from a Carbon wall to a metal wall has introduced changes in the L-H transition power threshold ($P_{LH}$) and in the value of the density at which the $P_{LH}$ reaches a minimum, $n_{e,min}$. New studies in SET would clarify to what extent the changes are due to the W wall or the W divertor, and if they are affected by boronization.

Due to limits in Tritium consumption, L-H transition experiments with Tritium in JET-ILW concentrated on the Horizontal Target configuration. A combination of various datasets show a clear shift in $n_{e,min}$ and $P_{LH,min}$, with lowest values for Tritium, followed by DT, D and H [80]. It is found that for each isotope a fixed fraction of the Greenwald density determines $n_{e,min}$.

Studies of H+T mixtures with different Tritium concentrations (and corresponding $A_{eff}$) at fixed density revealed that in fact the dependence of $P_{LH}$ on $A_{eff}$ is not linear [81] and critical kinetic profiles of $n_e$ and $T_e$ must be achieved before the transition takes place, confirming that it is the L-mode confinement and its isotope dependency that leads to the $P_{LH}$ non-linear dependency on Tritium concentration.

L-H isotope experiments in SET could investigate low T concentration H+T or D+T plasmas, as planned in ITER's low neutron production campaigns, documenting the impact of heating mix, W PFCs and Boronization on both $n_{e,min}$ and $P_{LH,min}$.

Further, it is clear that for easier access to low neutron production good H-mode it would be beneficial to investigate dynamic transitions in heated H+T mixtures: apply heating during the current ramp up to enter H-mode with lower $P_{LH}$ at lower $I_p$ and $n_e$ (as long as shine-through limits allow it [82]), raise power, current and density together just slowly enough to remain in H-mode. SET experiments characterizing plasma behaviour in these dynamic conditions, with appropriate isotopic content, could inform future ITER operation.

Considering that most future devices aim to use extrinsic impurities to alleviate divertor heat fluxes, it would also be important to study the impact of impurity seeding on $n_{e,min}$ and $P_{LH}$, for different isotopes. It is already known that impurity effects at JET are more complex than simply subtracting the core radiated power from $P_{LH}$ [79],[80].

### 5.2 Confinement scaling studies

Consider for instance the dimensionless scaling studies in between D and T in Vertical Target L-mode plasmas [83]. They were carried out in Vertical Target configuration, at 2.2T, 1.8 MA in D and 3T 2.4 MA in T, NBI heated. Matched profiles enabled characterisation of the energy thermal confinement time dependency on A as $\Omega_i \tau_{E,th} \sim A^{0.48}$, while the matching comparison between H and D displayed no mass scaling of $\tau_E$. A wealth of new studies could now be carried out, characterising confinement scaling in other configurations and with different heating and isotope mixes. Modulated ECRH would be an additional tool to investigate local transport, combined with modulated gas puff to investigate also the particle transport scaling.

On the other hand, dimensional similarity experiments in H-mode [84] focused on spanning H, D and T single isotope plasmas in comparable conditions, and in 50/50 DT mixtures. NBI heated ELMy H-mode plasmas with 1.7T, 1.4 MA in Corner configuration were compared. Variations of fuelling rates and the upper triangularity were used to attempt to match ELM frequencies across species. Different isotope mass scalings for the pedestal density were observed depending on the gas fuelling level. Increased W radiation in T plasmas was reported, it is possible this would be enhanced by the W wall. With the added ECRH capability, H-modes with higher $I_p$, $B_{tor}$ and core temperatures, closer to fusion-relevant conditions, in Tritium and DT plasmas could be investigated in SET, also enabling better edge density measurements by reflectometry.

Different combinations of heating systems would help elucidate the influence of rotation, $T_e/T_i$, $E_r$, fast ion content, $\nabla p$ and electrostatic vs. electromagnetic effects on plasma confinement. Recent theoretical studies show that various different mechanisms can play a role in the isotope dependencies of confinement [85],[86].

### 5.3 Other isotope studies

With the additional power and W PFCs, isotope studies of pedestal structure, ELM behaviour [87] and pedestal stability [88] carried out in DTE2 and DTE3 would be possible in SET at higher field and/or current, and in more configuration. The effect of the isotope mass on the pedestal and ELMs in low and high collisionality plasmas could be investigated.

Now that early ITER operation is expected to focus on low neutron production, studies of H-mode quality in Horizontal Target and Corner configurations with mixtures of H+T and D+T with low Tritium concentration would be extremely desirable, especially comparing the types of H-mode accessed with dominant ECRH *vs*. dominant NBI heating.

## 6. Fusion diagnostics

JET has state-of-the-art fusion diagnostics, including neutron camera, spectrometer, gamma ray detectors, neutral particle analysers, etc... This very complete set allows synergistic developments of essential new and improved fusion diagnostics. Some possibilities are described below.

### 6.1 Alpha particle diagnostics in tungsten devices by gamma-ray spectroscopy

Measuring the alpha particle phase space is essential to understand the physics of burning plasmas. It is experimentally challenging and will require the joint interpretation of measurements from various diagnostics. Gamma-ray spectroscopy is one of the few demonstrated methods that can access the alpha particle phase space by the observation of the line emission from spontaneous nuclear reactions between the alpha particles and impurities in the plasma [89], [90]. Gamma-ray spectroscopy at JET has relied on spontaneous nuclear reactions with $^{12}C$ or $^{9}Be$ impurities, associated to the respective plasma facing materials. In the Deuterium-Tritium DTE2 and DTE3 campaigns with a Be first wall, alpha particle measurements based on gamma-ray emission spectroscopy were demonstrated, both with a high purity germanium detector [40] and with a $LaBr_3(Ce)$ scintillator [91]. The 4439 keV line (nominal energy) from the reaction $^{9}Be(\alpha,n\gamma)^{12}C$ was measured, thanks to a 1% concentration of $^{9}Be$ impurities provided by the wall.

The re-baseline of the ITER project means that Be reactions are no longer expected, and alternative nuclear reactions must be sought to observe alpha particles using gamma-ray spectroscopy. The natural candidate in a boronized device is the $^{10}B(\alpha,p\gamma)^{13}C$ reaction. It has a smaller cross section compared to $^{9}Be(\alpha,n\gamma)^{12}C$ but it might provide enhanced alpha particle velocity space capabilities as it leads to three, rather than one, gamma-ray emission peaks.

A JET-based SET would begin already with a first-class suite of thoroughly tested gamma-ray diagnostics. SET is therefore the ideal platform to demonstrate and fully develop alpha particle diagnostic capabilities based on reactions with Boron in a variety of scenarios. These can encompass both D-$^{3}He$ plasmas, where alpha particle production is obtained via the $^{3}He(d,p)\alpha$ reaction, and reactor relevant DT plasmas, which pose more challenging measurement conditions due to the neutron induced background. Important parameters to be documented are the Boron levels required to enable measurements, the signal-to-background ratio depending on the scenario and the detailed relation between the experimental data (intensity and shapes of the relevant gamma-ray emission lines) and the properties of the alpha particles. At SET it could be experimentally established if the level of boron required for accurate alpha measurements leads to excessive Tritium retention.

SET would also enable a feasibility study of alpha particle diagnostics by alternative gamma-ray reactions that do not require boron as the target. This could support, for example, reactor operational scenarios that do not use boron, as may be desired to minimize Tritium retention. In this case, a possibility is to rely on $^{22}Ne$ as the target for the gamma-ray reaction. Neon may be injected also to study highly radiative detached divertor scenarios in SET.

### 6.2 Testing ITER relevant neutron spectrometer concepts for DT plasmas

Neutron spectroscopy measurements of DT plasmas are of paramount importance for the development of a fusion reactor. They can be used for the determination of the core ion temperature, the fuel ion ratio, the thermal/non-thermal ratio of the fusion power and the generation of supra-thermal fuel ions by the heating systems. The most successful neutron spectroscopy measurements of DT plasmas at JET have been based on two different detector concepts: the magnetic proton recoil (MPR) neutron spectrometer [91],[92], and diamond detectors [93].

At ITER, neutron spectroscopy measurements will be based on a suite of four instruments, called the high resolution neutron spectrometer (HRNS) [94]. For Deuterium plasmas, a (forward) time of flight spectrometer will be used, based on the successful experience with TOFOR at JET. For Deuterium-Tritium plasmas, instead, measurements will be enabled by a thin foil proton recoil (TPR) instrument, a diamond detector and a backward time of flight (back-TOF). No MPR detector will be developed for ITER, due to space limitations. Of the three HRNS instruments for the DT phase of ITER, the back-TOF has never been built and tested in a DT tokamak, which poses some risks for ITER. A back-TOF could be developed and tested at the SET, for example by placing it in the same position as the TOFOR spectrometer, which cannot be used in Deuterium-Tritium plasmas. Again, SET would be the ideal tokamak to test the back-TOF, as measurements with this detector technique could be directly compared with data from the MPR and diamond detectors.

### 6.3 Fusion power measurements by gamma-ray spectroscopy

Fusion power measurements are essential for the licensing of a DT reactor, and to assess plasma performance. The ITER nuclear regulator requires frequent measurements of fusion power with at least two independent methods, both with at least 10% accuracy. This is likely to be required by any regulator of any fusion device that uses DT mixtures. Therefore, qualifying new fusion power measurements would be a very substantial contribution of the SET to the fusion roadmap.

The primary method for measuring fusion power is by 14 MeV neutron counting, as done with the JET neutron cameras [95]. Although an accuracy in the range 7-10% has been demonstrated with this method, absolute neutron counting is time consuming as it requires extensive pre-operation calibrations. A new technique has been demonstrated at JET in the recent DT campaigns, relating the fusion power to the emission of 17 MeV gamma-rays from the fusion reactions [96], [97]. For the first time in a tokamak, the relative probability that a 17 gamma-ray is emitted by the DT fusion reaction with respect to neutron production (branching ratio) was determined and the spectral shape of the emission was measured directly. Despite being a success, the measurement could be demonstrated only at relatively low neutron rates in the range from some $10^{16}$ n/s to some $10^{17}$ n/s, due to the non-optimized neutron and gamma-ray shielding of the detector used for this first-time result. Moreover, neutron measurements, which are necessary to determine the branching ratio, were also not available on the same line of sight as the gamma-ray measurements and thus it was not possible to achieve the desired accuracy of 10% for the method to be fully ready for future fusion devices. SET could contribute to advancing the method to the level required for a fusion power reactor. By integrating a diamond neutron spectrometer along the line of sight of the gamma-ray detector, the uncertainty on the determination of the branching ratio could be reduced to the level required to achieve an overall 10% accuracy on the determination of the fusion power. It may be possible to demonstrate fusion power measurements up to the highest fusion power levels that can be achieved by DT plasmas in SET, corresponding to neutron rates in the range of some $10^{19}$ n/s, i.e. about two orders of magnitude higher than what has been achieved so far at JET.

## 7. Fusion nuclear technology, impact on materials, and safety

Once appropriate DT plasmas in SET are developed, many fusion technology issues [15] could be addressed, some with existing JET systems, others with modifications or upgrades. As SET is much smaller than ITER, its maximum first wall neutron flux is only one order of magnitude smaller than that of ITER. The accumulated neutron fluence during DTE2 and DTE3 at JET was at the same level as that expected at a rear ITER port plug at the end of the machine lifetime, or as that expected behind the blanket at the end of the first ITER DT phase [16]. Therefore SET, with enhanced neutron production in ECRH-aided DT campaigns, can uniquely provide new tests for various fusion nuclear and materials issues.

Based on the summary presented in [15][16], we briefly present in what follows some of the many nuclear studies that could be addressed in SET.

### 7.1 Water activation Studies:

Activation of cooling water is a significant concern for radiological safety. 14 MeV fusion neutrons activate water via the reaction $^{16}O(n,p)^{16}N$ while 2.45 neutrons MeV do not [98]. Measurements of cooling water activation by 14 MeV neutrons with improved instrumentation would enable better validation of neutron transport & materials activation codes. Initial attempts in DTE3 were hampered by sub-optimal installations. A minor upgrade, already planned but not executed in DTE3, is available.

### 7.2 Activated Corrosion Product studies

Corrosion, erosion, and dissolution of materials in cooling circuits are concerns as they can lead to the mobilization of corrosion products into the fluid, which, under neutron irradiation, become Activated Corrosion Products (ACPs). ACPs represent a significant radiological hazard in high performance machines, as they are transported by the coolant to areas accessible to personnel for maintenance. The quantification of their impact in ITER is crucial for assessing the radiological hazard. Estimates of Occupational Radiation Exposure need validated computational tools to enable safe waste management, and maintenance planning. A proper loop and diagnostic could be designed and installed at SET. This would be a major enhancement, for which a first feasibility study is available [99]. Alternatively, appropriate diagnostic systems need to be installed to measure the mass/activity of the ACP (medium enhancement but pre-analysis needed). The results of these measurements would enable validation of the appropriate safety codes, and evaluation of feasibility, safety case and cost of sampling cooling loops in future fusion devices such as ITER or DEMO.

### 7.3 Single Event Effect studies on electronics and verification of shield effectiveness

The interaction of single neutrons with components used in electronic circuits can corrupt signals and/or introduce errors in data and control systems. Initial exposure studies in WEST with D-D neutrons [100] were continued by exposing the same electronic components to the DTE3 D-T neutrons, but in suboptimal conditions, in the JET basement. It would be necessary to move the existing electronics test bench inside the SET to the Torus Hall to complement the earlier studies. We might see, compared to DTE3, increased effects of high energy neutrons and it would be possible to evaluate the effectiveness of proper electronic shield cabinets.

### 7.4 Short-term activation study of ITER /DEMO materials

Real ITER material samples can be exposed in the Irradiation End to characterize the neutron-induced radioactivity and validate data for the prediction of ITER materials activation. A few samples were exposed during past JET-ILW DT campaigns [101]. Samples irradiated in long-term irradiation stations during the previous DD, DT, and TT campaigns were retrieved after several months, thus the short-term nuclides had already decayed and no information from decay gamma spectra could be obtained. Carefully planned exposure and extraction of samples immediately after a SET DT campaign would provide invaluable data on short term activation.

### 7.5 Photo-neutron production from run-away electrons in W/B

Beryllium tiles exposed to intense gamma fluxes produce photo-neutrons via $^{9}Be(\gamma,n)^{8}Be$ (1.66 MeV energy threshold). The photo-neutrons generated by prompt gammas during operations were negligible compared to the DT plasma neutron emission, but at shutdown the decay gammas from neutron activation of materials does generate a delayed photo-neutron source that can increase the dose and activate remote handling and transportation equipment.

Experimental evidence of this process has been revealed during the JET shutdown following DTE2. With adequate diagnostic enhancements, these processes could be characterised in the W PFC + boronization environment envisaged in ITER. Experiments could start in Deuterium campaigns in SET.

For nuclear issues such as those described above, the development and validation of measurement tools and associated codes are equally important, and SET's DT capabilities prove to be essential for most of these studies. We won't review the complex code and licensing needs here, also described in [15].

## 8. Summary, discussion and conclusions:

In this perspectives article we have introduced a possible DT-capable tokamak with W PFCs, a boronization system, 10 MW of ECRH and suitably enhanced diagnostics, based on JET, which we call SET for short. The impact of W on operating conditions was discussed, based on available information and ITER's concerns. Options for development and operation of next-step ECRH systems in a fully integrated tokamak facility were presented. DT plasmas to be expected in SET were described, illustrating how existing operating regimes could be made much more fusion relevant thanks to the addition of electron heating which would contribute to equalising $T_e$ and $T_i$, as well as raising core $T_i$ and nuclear output. It would therefore be possible to investigate multiple fusion-relevant issues, including isotope effects. Notably, new operating scenarios with increased neutron production would enable necessary testing and validation of fusion and nuclear diagnostics at SET. This could facilitate regulatory approval and DT operation for next step devices, including ITER.

Mitigating **operational** risks for re-baselined ITER and other next step fusion devices not only necessitates targeted developments in DT experiments, but also training of the next generation of fusion scientists, technicians, and engineers. Nuclear expertise in the Tritium technology field is scarce and will fade if not nurtured in integrated nuclear tokamak facilities such as SET.

SET would be an ideal vehicle to maintain momentum towards the goal of producing fusion energy and accelerate DT operations in next step devices, certainly in view of the ITER delays announced recently [1],[2],[3],[4].

There is a need for a range of fusion facilities to advance the field. The research programme we described here is complementary to that of other existing, new and planned facilities. For instance, SET doesn't have superconducting coils, and thus would not compete with facilities that aim to study fully non-inductive steady-state operation, such as JT-60SA [102]. For fusion material studies, facilities such as LIBRTI (UK) and IFMIF-DONES (EU) [104] remain a necessity for the fusion programme, as candidate reactor materials need to be tested and qualified for use in the harsh environment of a fusion reactor. Operating in DT and producing 14 MeV neutrons, SET would have a much lower neutron fluence than what is expected in fusion power plants, but would still enable relevant tests, especially for validation of nuclear simulation tools. Tritium breeding studies are essential for fusion development, but we don't propose to do them in SET. One might consider an alternate, less ambitious, upgrade scenario, concentrating on what is a truly unique capability of SET: DT operation producing fusion-relevant plasmas. In this scenario, an ECRH plant would be built early on, and gyrotrons added as soon as they became available. A large variety of fusion devices are now studying W walls and boronization, specialists in the relevant subjects would need to evaluate the potential contributions of W PFCs in SET and how they compare to expected findings elsewhere. On the other hand, the need for additional electron heating to optimise JET DT capabilities has long been acknowledged by the fusion community.

There is an important clarification that needs to be made: JET is not old, it is a mature device that delivered sterling performance just before it was shut down at the end of 2023. JET could become SET at a fraction of the cost of a new facility with equivalent capabilities. JET's main components are at half of their expected lifetime. In 2023 the JET Team, now with the experience acquired in DTE2, was able to produce almost 4000 pulses between the start and end of the experimental campaigns, as shown in Fig. 7. This compares very favourably with the previous operation years, which typically delivered between 1000 and 3000 pulses per year. Detailed analysis of JET performance can be found in [107] and [108].

Although JET repurposing and de-commissioning started in January 2024, the facility has not been destroyed yet and there is enough Tritium available for a new DT campaign in a renewed SET. See Appendix C for a conceptual description of the hardware changes to be implemented. Hands on expertise of DT operation at SET would be invaluable to future fusion developments. For background on how ITER design decisions have affected the JET facilities, see Appendix D.

The EU fusion community is working towards an integrated fusion strategy. A recent report of the Fusion Expert Group [116] highlights the need for an assessment of the gaps in the EU fusion programme, and the importance of international collaboration to bridge them. The report also highlights the importance of maintaining know-how and developing a workforce strategy. In that respect, we would add that SET could provide focus to the EU+UK fusion programme, maintain fusion know-how and train the next generation of fusion experts in a real machine.

We trust the research topics we highlighted here would be useful to evaluate the benefits of a device such as SET to the fusion programme. Note that the discussion we present in this article isn't exhaustive: additional issues might be proposed by other experts with different interests. The research programme described here can most easily and timely be implemented in SET, but it may be also relevant for other DT-capable devices.

In conclusion: SET, a DT-capable tokamak, could make unique contributions to the EU fusion programme. These include a timely investigation of the impact on operation of all-W PFCs, qualifying the need for boronization and the associated management of fuel retention; an evaluation of the impact of additional electron heating provided by ECRH to develop scenarios that enable unique studies in plasmas with Tritium, and especially DT, including studies of thermonuclear alphas and isotope studies. Notably, SET would enable further development of fusion diagnostics and nuclear technology tools. It could contribute to accelerate fusion development, so needed to decarbonise the world's energy supply.

## 9. Acknowledgements:

The author would like to thank D. Alegre, E. de la Cal, J. Deane, M. Henderson, I. Jepu, J. García, L. Garzotti, D. King, E. Lerche, A. Loarte, M. Mantsinen, M. Maslov, M. Nocente, M. O'Mullane, J. Ongena, M. Siccinio, K.D. Zastrow, and contributors that prefer to remain anonymous, for crucial input and discussions. Like the named author, all provided information and contributions as individual fusion experts, not representing institutional views.

This article is an extended version of the presentation made at the 50th EPS Conference on Contr. Fusion and Plasma Phys, 8-12 July 2024, in Salamanca, Spain, [105]. The accompanying poster received the People's Choice Award, selected by the conference participants [106]. We are also grateful to the editorial team of Fundamental Plasma Physics, for their kind invitation to submit this manuscript.

Most of all, we express our thanks to the whole JET Team. It has been a privilege to work at JET with such a dedicated multinational team of experts. The support of all those that signed the petition to extend JET beyond 2023, https://www.petitions.net/petition_to_extend_jet, provided the inspiration for this article. We are grateful to the JET Joint Undertaking, EFDA, EUROfusion and UKAEA for jointly funding fusion research at JET.

This work was supported in part by grant PID2021-127727OB-I00, funded by the Spanish MCIN/AEI/10.13039/501100011033 and by ERDF "A way of making Europe".

**Appendices: background and technical details**

**Appendix A: Tungsten Wall**

The new wall can be installed by Remote Handling, in a similar manner to the installation of the JET ITER-like Wall (JET-ILW) [109]. At the time, the wall change took 18 months of shutdown, constituting a great training experience for the Remote Handling group.

Fig. 1 illustrates the JET-ILW wall, and helps explain the proposed changes. The Be limiter tiles, green coloured, would be replaced with W-covered tiles (CFC blanks were still available on-site at the time of writing). Clean new tiles would be installed in the inner wall (blue in the figure), without Be. Options for remote areas of the inner wall are W-coated CFCs, W-coated Inconel [111], W-coated stainless steel (Eurofer or ODS steels if available, as they are the main option for the main wall at DEMO [112], or simply naked Inconel or steel tiles. Damaged or Be-covered divertor tiles would need to be replaced (this is a concern on the tile 1 row in the high field side [110]).

**Appendix B: ECRH**

**Option 1: 170 GHz system**

The addition of an electron cyclotron resonance heating and current drive system at JET has been planned twice in the past. An initial design was developed in 2001-2003, [113], but it wasn't approved. The project was costed and approved by the European fusion authorities, but somewhat ironically it was decided to install an "ITER-like" Be wall and W divertor at JET instead, the JET-ILW project, now rendered non-ITER-like by the ITER re-baselining.

In 2011 a new project for ECRH installation was developed [30],[32], seeking synergies with the ITER ECRH system [33]. The ECRH system was envisioned to enable JET to more closely mimic the then-expected ITER-like plasmas, by increasing at JET the available electron heating, sawtooth control, neoclassical tearing mode control for access to high beta regimes, and current profile control to access and maintain advanced plasma scenarios and Internal Transport Barriers. Extensive simulations were carried out to optimise gyrotron selection and antennae design, based on known JET-C ELMy H-mode and, Hybrid and Advanced Tokamak plasmas best documented at the time. The design of an ECRH system with a set of 170 GHz gyrotrons would operate in X-mode 2$^{nd}$ harmonic at 3T, it is ready to be implemented with a predicted time-scale of 5 years, and remains suitable for the type of plasmas considered to be of interest then, typically 2.8 Tesla plasmas. The realization of the project was predicated on international collaboration, but it didn't obtain support from the ITER team.

**Option 2: multiple frequency systems**

A single frequency ECRH system can be rather limiting, basically fixing the toroidal fields with available ECRH. This is why more complex systems are under consideration for future devices.

Multi-frequency / multi-purpose systems are being designed with gyrotrons capable of operating with multiple different frequencies. For instance, based on the ITER choice, a single gyrotron could deliver at 136, 170, 204 or 238 GHz [36]. At these frequencies, the single-disk CVD diamond ITER output window is transparent for the microwaves due to its resonances at $n/2 \times$ wavelength in the dielectric.

Another option are gyrotrons with frequency step-tunability, during a plasma discharge. The tunable range is ±10 GHz in steps of 2 – 3 GHz around a center frequency of e.g. either 136 or 170 or 204 or 240 GHz. This kind of operation requires a new type of output window, preferably a Brewster-angle window, and a fast-tunable gyrotron superconducting magnet.

Integrated testing of such systems at SET, would potentially benefit ITER, STEP [34],[35] and DEMO [36], while enabling more flexibility in terms of the toroidal field optimization for ICRH and ECRH in SET.

Based on STEP estimates, it's expected that, from project start, the first 4-5 MW could be available in 5 years, followed by additional 2-4 MW per year to be delivered in the following years. If work started in 2026, 10 MW could be ready by 2033.

**Appendix C: From JET to SET?**

This article presents a scientific assessment of the value a DT-capable tokamak such as SET might bring to fusion research. We deliberately avoid policy and political discussions, such as costing, priorities or funding. Still, we must consider the practicalities of the following questions: assuming additional funding and suitable cooperation agreements materialise, would it be possible to transition from JET in its present state to SET? What would be required, in terms of hardware change? How long would it take?

The on-going JET decommissioning activities have by now impacted some of the JET infrastructure, potentially increasing the time it might take to re-start operations. The UKAEA also has a re-purposing plan by which buildings that house JET infrastructure are being assigned to other uses within the fast-developing UK's fusion strategy. Changing such plans would likely not be trivial, and entail at least short-term costs.

**Assuming** a joint EU-UK decision to transition from JET to SET (presently not being contemplated), funding and swift re-staffing, it is estimated unmodified JET could be ready for restart 2 years later, after new spares and equipment are purchased to replace already discarded, removed or outdated elements (notably control elements of the power infrastructure, affecting toroidal field and auxiliary heating control and electrical connections, some diagnostics). It is widely believed that there is enough Tritium left on-site for future DT campaigns, but concrete information hasn't been released.

The transition from JET to SET could most effectively take place with design of new elements being carried out in parallel with refurbishment activities and recovery of operations in Deuterium. We have not discussed desirable Deuterium experiments in this article, but certainly there is no shortage of ready-to-go important experimental proposals that could not be executed or finalised in the last years of JET operation.

The first steps would include inspection, refurbishment and maintenance of the existing infrastructure, such as the Active Gas Handling System, cryogenics and cooling plant (replace/repair elements, add He liquefiers), baking plant, Glow Discharge Cleaning electrodes, power supplies, buildings, etc... A review of the neutron budget and the Safety Case, redefining the neutron limit, would need to be undertaken.

The wall change, as described in Appendix A, could be based on the JET-ILW project [109], which required an 18-month shutdown. The initial idea is to replace Be tiles with W-coated CFC tiles, and install clean metal tiles in the inner wall. W-coated graphite tiles may also need to be replaced, especially on the tile 1 row in the high field side, were some of the W coating is damaged [110]. New tiles would need to be installed to replace tiles already removed for retention studies.

Two options for ECRH and ECCD systems are discussed in Appendix B. If started in 2026, it is estimated 10 MW could be ready by 2033. As proposed in [32][30], installation of the ECRH system in Octant 2 would entail the removal of the ICRH ITER-like (ILA) antenna, a system no longer operational. Protection from microwave radiation would be required for various diagnostics. A new building would be needed for the gyrotrons and associated power supplies.

JET already has integrated control systems. For SET it would be wise to undertake upgrade studies for data acquisition (especially real-time) and control systems. These would include control loops for wall & divertor protection, disruption avoidance & disruption mitigation, control of heat deposition, q-profile, ITB location, MHD, NTM, detachment, dud detection, termination, runaways, etc…, with multiple actuators. ITER-like data storage and control architecture could be tested, including various potential applications of AI.

The elements of the research programme presented in this article would need to be evaluated and costed in detail by experts. For now, we suggest the following SET development timeline could be considered:

- 2026-2028 refurbishment and minor upgrades, conceptual design of major upgrades
- 2029-2030 begin staged installation of upgrades, re-start in Deuterium.
- 2031-2032 experimental campaigns in Deuterium, installation of ex-vessel upgrades.
- 2033-2034 install in-vessel upgrades, change wall, full ECRH.
- from 2034 carry out experiments in D, H+T, T and D+T, including Tritium-poor mixtures to mimic ITER's low neutron production campaign and for specific isotope studies.

## Appendix D: On JET and ITER decisions

A brief summary about JET and ITER re-baseline decisions is presented here for those unfamiliar with the history.

In the 2016 ITER Baseline Research Plan, ITER had a Beryllium wall and a Tungsten divertor, and the JET-ILW project was carried out to prepare ITER operation.

The EUROfusion Roadmap, published in 2018 [115], was built on the expectation of observing a First Plasma in ITER in 2025-2026. A footnote in page 8 of [115] explains the expectations for JET operation then: "The end date of JET operation was under discussion at time of writing. There are strong arguments to keep JET in operation as close to the first plasmas on ITER as possible". Unfortunately, in 2023 it became clear that first plasmas would not be seen at ITER in 2025-2026, and the ITER Team began to develop an alternate plan, but somehow support for the view that JET operation should therefore continue disappeared.

During 2023 the ITER team, working closely with Asdex collaborators, developed the alternate plan, now widely known as the ITER re-baselining. It was first discussed publicly at the IAEA FEC 2023 conference [114]. A simplified overview of the ITER Re-Baseline plan was given in the introduction, see Table 1.

It is now clear that no DT fusion plasmas are expected in ITER before late 2041 or 2042. SET could act as keeper of the knowledge of DT operation in tokamaks until then, advance physics understanding, develop operating regimes and develop and validate fusion and nuclear diagnostics.

**Figure captions:**

Figure 1: From JET-ILW to SET ITER-relevant W wall: Beryllium limiter tiles (green) to be replaced with W-covered tiles, Be-coated Inconel tiles in remote areas (blue) to be replaced with naked Inconel or Eurofer or W-coated metal tiles.
Figure reproduced from G Arnoux et al, 2014, Phys. Scr. 2014 014009, with permission.

Figure. 2: Dashed blue lines depict the 2023 Tritium-rich JET Pulse Number (JPN) 104522, red lines correspond to adding 10 MW of ECRH to the same target, as described in the text.

Figure 3: Equipartition time in terms of $n_e$ and $T_e$ in a 50/50 D-T plasma. Shaded in pink the area relevant to the plasmas being studied.

Figure. 4a) Power deposition after Coulomb collisional redistribution of the ICRH & NBI heated JPN 99948 Baseline DT pulse, and additional curves showing simulated power depositions when 3MW of central ECRH is added. 4b) Evolution of the temperature as a function of time when adding 3MW of EC power to JPN 99948, at different points in time, showing that conditions can change from $T_i$=$T_e$ to $T_e$>$T_i$

Figure 5: a) predicted electron density and b) electron and ion temperatures (dashed and solid lines, respectively) for Deuterium shot JPN 96730. The black lines are the reference profiles without any ECRH power, the red and blue lines correspond to simulations where 5 and 10 MW of ECRH were added on top of the ICRH and NBI power used in the experiment.

Figure 6: Typical plasma shapes used in various studies at JET-ILW (a) Corner (CC), (b) VT, Vertical Target (VT) and (c) Horizontal Target (HT).

Figure 7: JET-ILW pulses per year as measure of availability.

**Figure 1**

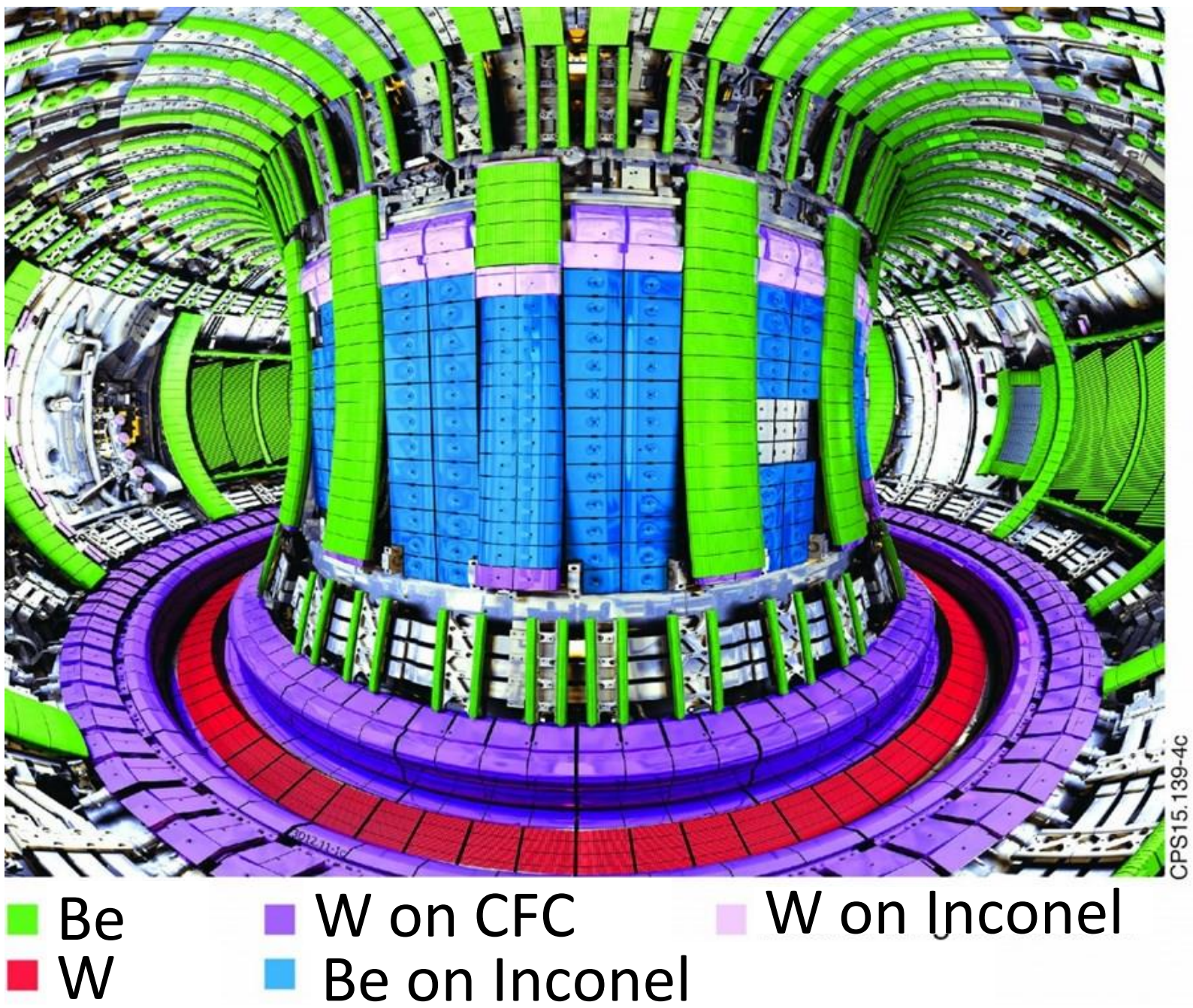

Fig. 1: From JET-ILW to SET ITER-relevant W wall: Beryllium tiles (green) to be replaced with W-covered CFC tiles, Be-coated Inconel tiles (blue) to be replaced with naked Inconel or Eurofer or W-coated metal tiles in remote areas.
Figure reproduced from G Arnoux et al, 2014, Phys. Scr. 2014 014009, with permission.

**Figure 2**

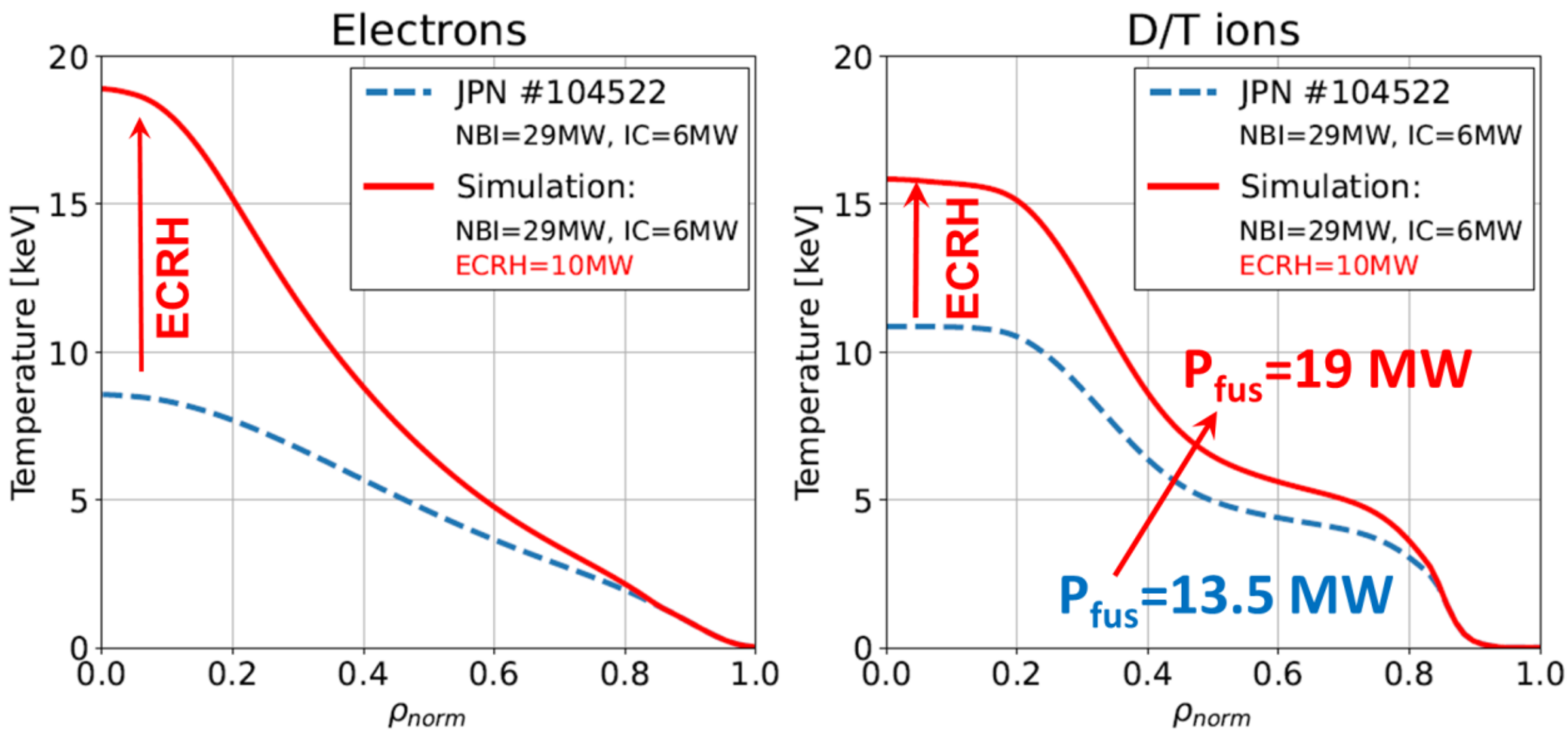

Fig. 2: Dashed blue lines depict the 2023 Tritium-rich shot JPN 104522, red lines correspond to adding 10 MW of ECRH to the same target, as described in the text.

**Figure 3**

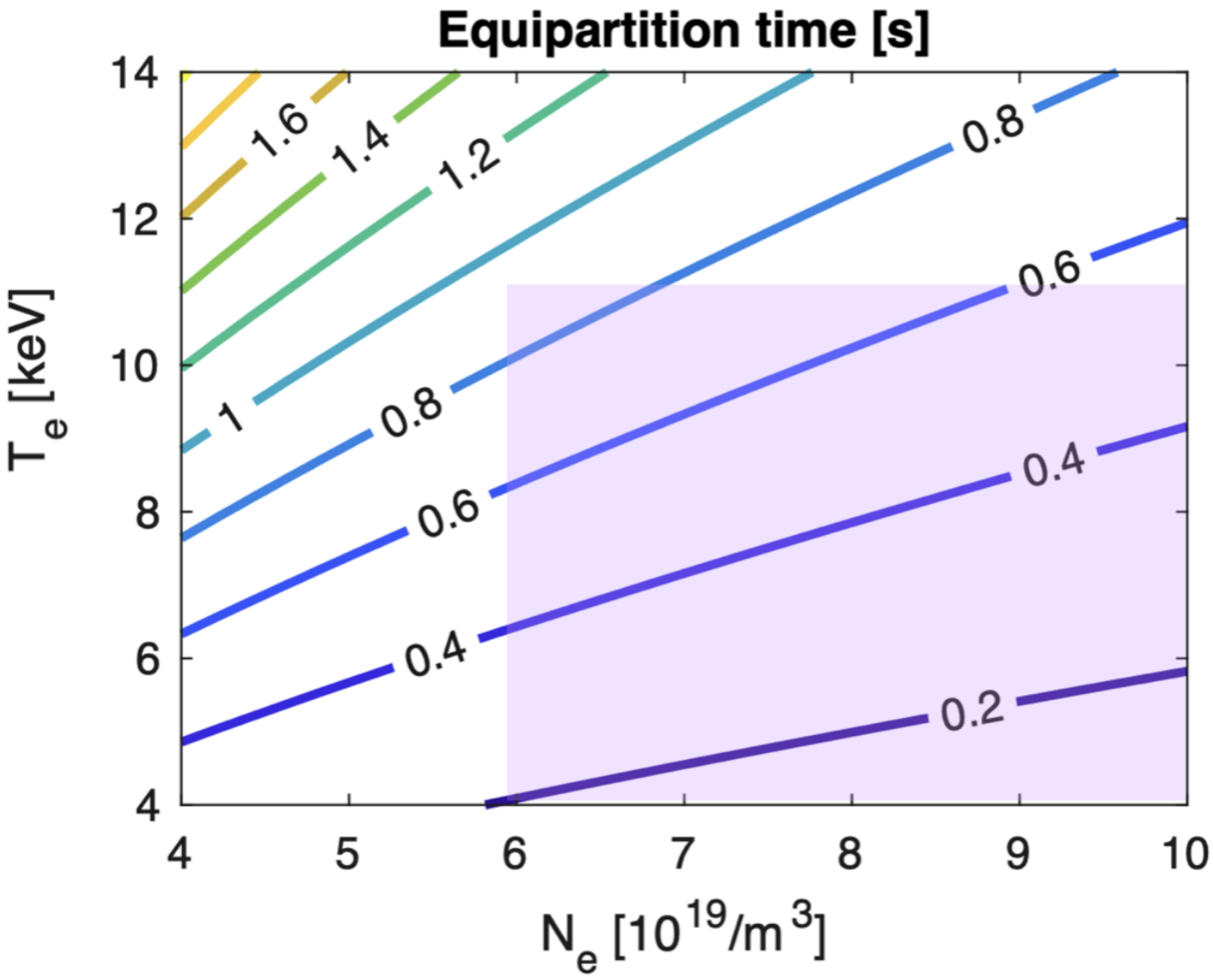

Fig. 3: Equipartition time in terms of $n_e$ and $T_e$ in a 50/50 D-T plasma. Shaded in lavender is the area relevant to the baseline plasmas in JET-ILW.

**Figure 4**

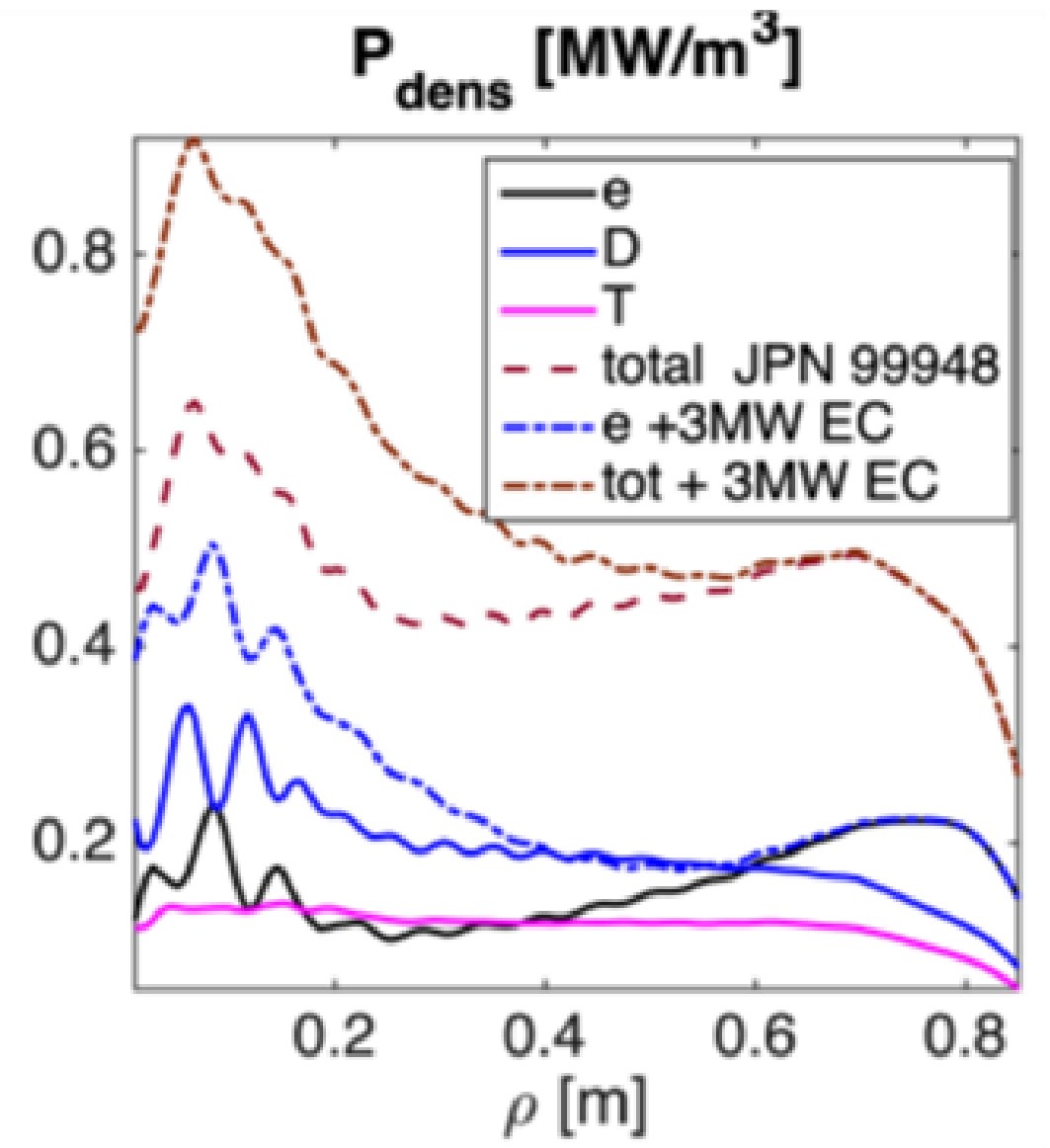

Fig. 4a: Power deposition after Coulomb collisional redistribution of the RF & NBI heated JPN 99948, and additional curves showing simulated power depositions when 3MW of central ECRH is added.

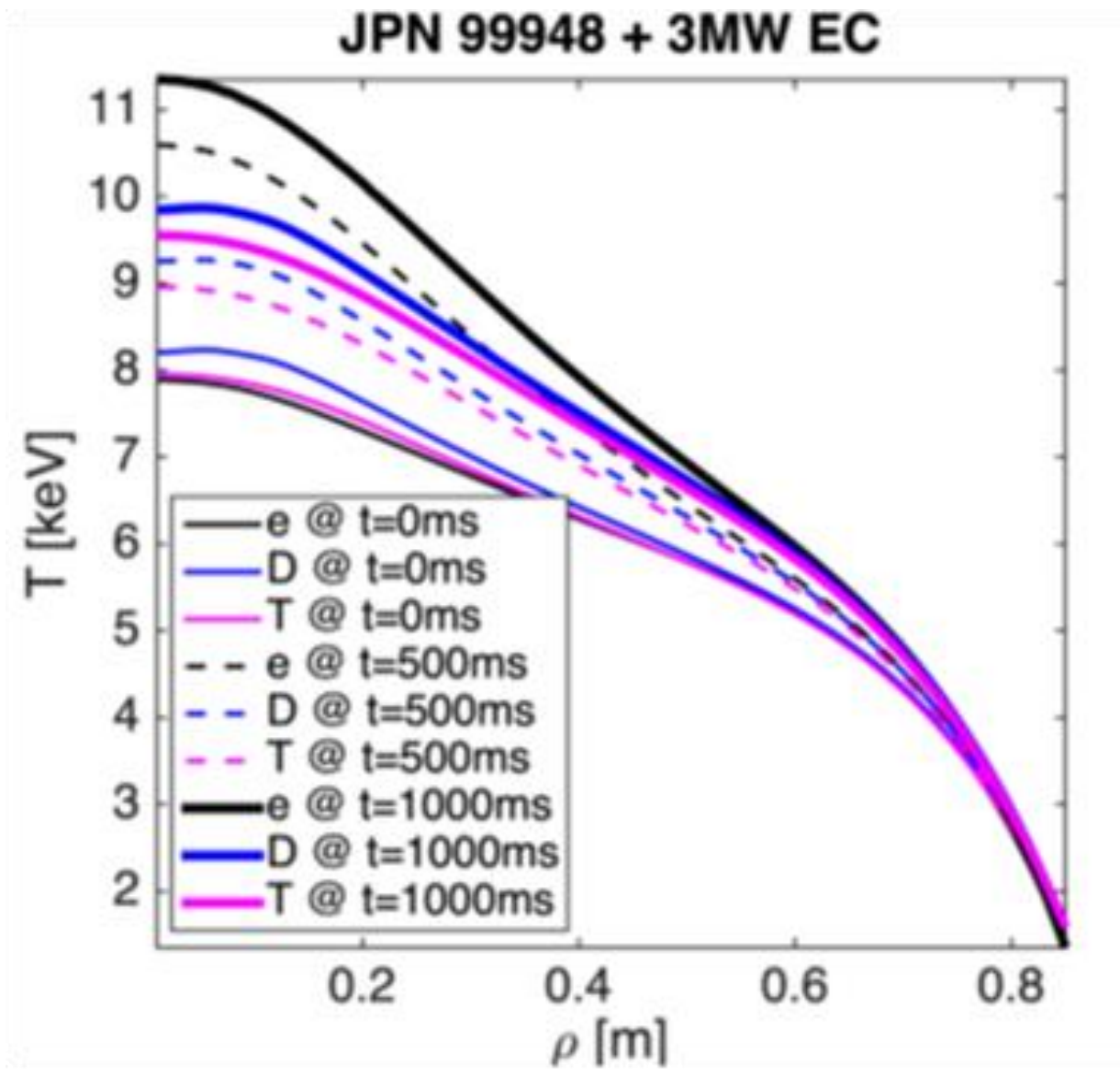

Fig. 4b: Evolution of the temperature as a function of time when adding 3MW of EC power to JPN 99948, at different points in time, showing that conditions can change from $T_i$=$T_e$ to $T_e$>$T_i$

**Figure 5**

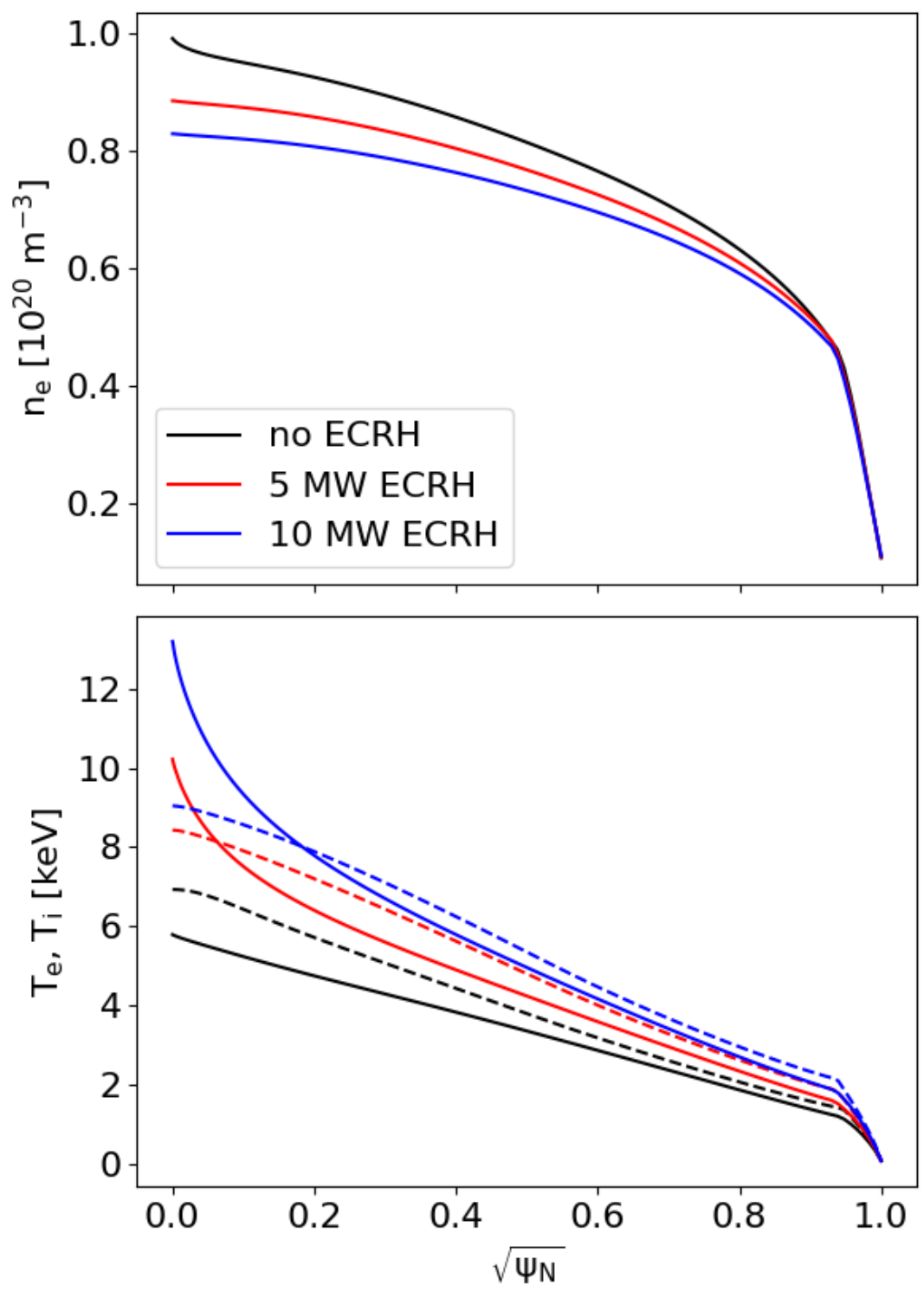

Fig. 5: a) predicted electron density and b) electron and ion temperatures (dashed and solid lines, respectively) for Deuterium shot JPN 96730. The black lines are the reference profiles without any ECRH power, the red and blue lines correspond to simulations where 5 and 10 MW of ECRH were added on top of the ICRH and NBI power used in the experiment.

**Figure 6**

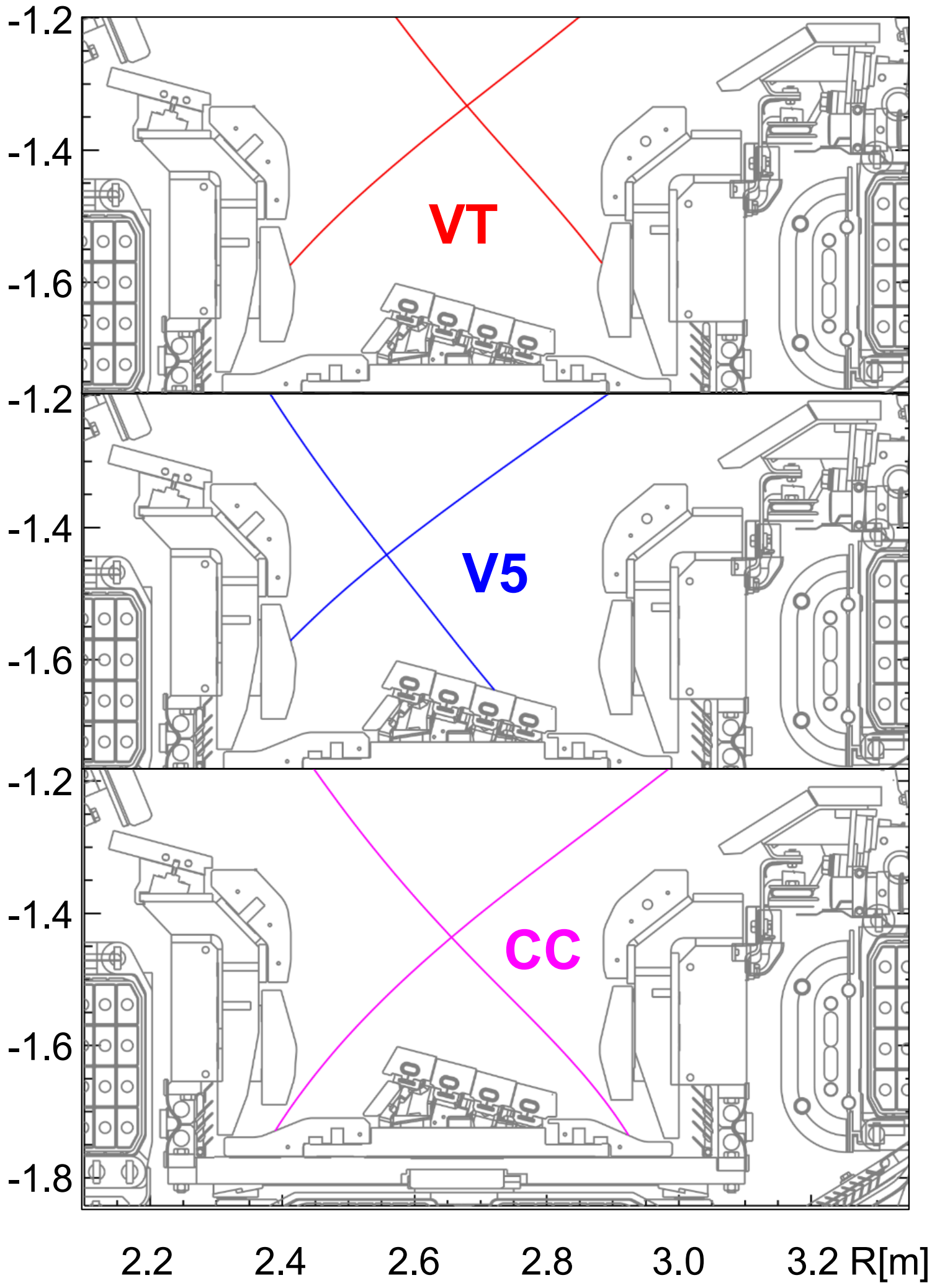

Figure 6: Typical plasma shapes used in various studies at JET-ILW (a) Corner (CC), (b) VT, Vertical Target (VT) and (c) Horizontal Target (HT).

**Figure 7**

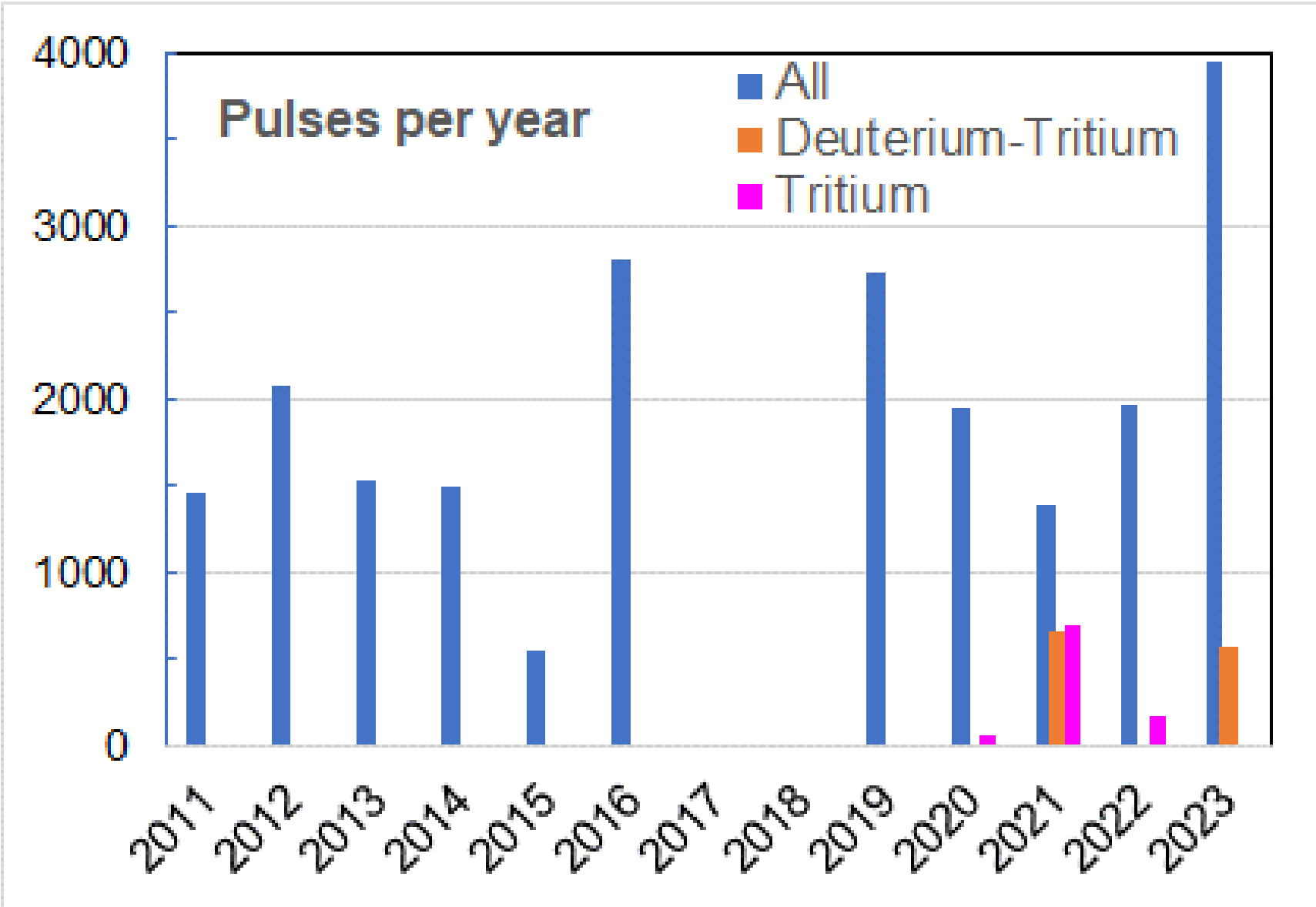

Figure. 7: JET pulses per year as measure of availability.